# High-temperature deep-level transient spectroscopy system for defect studies in wide-bandgap semiconductors

S. Majdi[1,*], M. Gabrysch[1], N. Suntornwipat[1], F. Burmeister[1], R. Jonsson[1], K. K. Kovi[1,2] and A. Hallén[3,4]

[1] Division for Electricity Research, Uppsala University, Box 534, SE-751 21, Uppsala, Sweden.

[2] Center for Nanoscale Materials, Argonne National Laboratory, Argonne, IL-60439, United States.

[3] Royal Institute of Technology, KTH-ICT, Electrum 229 SE-164 40, Stockholm, Sweden.

[4] Department of Physics and Astronomy, Ion Physics, Uppsala University, Box 516, 751 20 Uppsala, Sweden.

**Abstract**

Full investigation of deep defect states and impurities in wide-bandgap materials by employing commercial transient capacitance spectroscopy is a challenge, demanding very high temperatures. Therefore, a high-temperature deep-level transient spectroscopy (HT-DLTS) system was developed for measurements up to 1100 K. The upper limit of the temperature range allows for the study of deep defects and trap centers in the bandgap, deeper than previously reported by DLTS characterization in any material. Performance of the system was tested by conducting measurements on the well-known intrinsic defects in *n*-type 4H-SiC in the temperature range 300-950 K. Experimental observations performed on 4H-SiC Schottky diodes were in good agreement with the literatures. However, the DLTS measurements were restricted by the operation and quality of the electrodes.

* Corresponding author: E-mail: Saman.Majdi@angstrom.uu.se
Tel: +46 (0) 18 471 5854, Fax: +46 (0) 18 471 5810

**Introduction**

Substitution of chemically different impurity atoms into a host semiconductor can generate deep-level states in the bandgap. Additionally, substantial variation from the ordered lattice, for instance a vacant lattice position, can give rise to allowed states between the conduction and valence band close to the mid-gap. Such centers are not always ionized at room-temperature. Deep states act as efficient electron-hole recombination centers that reduce carrier lifetimes, particularly in indirect semiconductors. In some instances, deep-levels are deliberately introduced to increase switching speed for semiconductor devices by reducing the minority carrier lifetime, e.g. in fast switching power devices.

Deep-level transient spectroscopy (DLTS) is a powerful electrical characterization method based on measuring the depletion region capacitance of a Schottky or *pn*-junction[1–4]. If there are deep-levels present, this capacitance will change as a function of time due to the emission of trapped carriers[5]. It is complementary to other methods for identifying defects, e.g. photoluminescence and electron spin resonance. DLTS, however, being an all-electric method, gives information about electrically active defects and impurities and is a highly sensitive technique, capable of detecting and quantifying very low defect concentrations.[2]

Power electronics based on low-loss semiconductor devices are expected to play a leading role in the development of future power generation, transmission and distribution systems. Wide-bandgap (WBG) semiconductor materials, such as SiC, GaN, $Ga_2O_3$ and diamond are considered to be extremely promising. SiC is the most matured WBG material which has been frequently investigated using DLTS during decades, but still reveals defect related findings[6]. $Ga_2O_3$ is another interesting material which recently has attracted great attention[7,8]. In the case of diamond, the excellent material properties, for instance high breakdown field, high carrier mobility and very high thermal conductivity are even more promising. For electronic device applications, the quality of single-crystalline (SC) diamond has now reached a stage where it is possible to develop reliable devices[9–12]. In addition, diamond contains exciting impurities such as nitrogen vacancies (N-V), silicon vacancies (S-V) and silicon-hydrogen complexes. These impurities can be manipulated for e.g. room-temperature quantum information processing or quantum computation and cryptography[13–17].

Indeed, the study of deep defects in WBG semiconductors is of fundamental importance to understand how they affect the electrical properties by trapping/scattering carriers and up to what level the fabrication processes, etching and surface passivation are effected. Classical DLTS on for instance Si and GaAs need typically cryogenic temperature. The introduction of WBG semiconductors has however made it necessary to extend the temperature range. Hence, a unique high-temperature (HT) DLTS system (up to 1100 K) has been designed and developed for electrical characterization of defects and impurities in WBG materials. The system will fulfill the requirement of providing sufficient thermal energy to study mid-gap defects in WBG semiconductors such as SiC, ZnO, GaN and in particular $Ga_2O_3$ ($E_g$~4.8 eV) and diamond ($E_g$~5.45 eV).

**Instrumentation**

The HT-DLTS system is depicted in Fig. 1 & 2. To avoid oxidation of the samples and their contacts at high temperatures, the sample holder is situated inside a vacuum chamber connected to a turbo pump providing required environment. The samples, typically Schottky barrier diodes (SBDs) or *pn*-diodes, are mounted on an electrically conductive thin Tantalum (Ta) plate which functions as back contact. The plate, is placed on a rectangular Boralectric ceramic element, forming a homogenous circular (25 mm in diameter) heating area. A Boralectric heater is a mixture of a dielectric ceramic material (Pyrolytic Boron Nitride) and an electrical conductive material (Pyrolytic Graphite). This was chosen because of the demanding requirements concerning vacuum compatibility, high working temperatures and chemical inertness of the material. To prevent unnecessary heating of the chamber walls and the mounting stage in the chamber and also to assure even sample heating, a heat shield, consisting of two Molybdenum (Mo) shielding boxes and ceramic insulators were used. Two high temperature thermocouples (type K) are connected to the ceramic heater and the Ta plate directly under the sample. A Tectra HC3500 Controller/Power supply together with a Yudian AI 518P PID temperature-controller is used as power source.

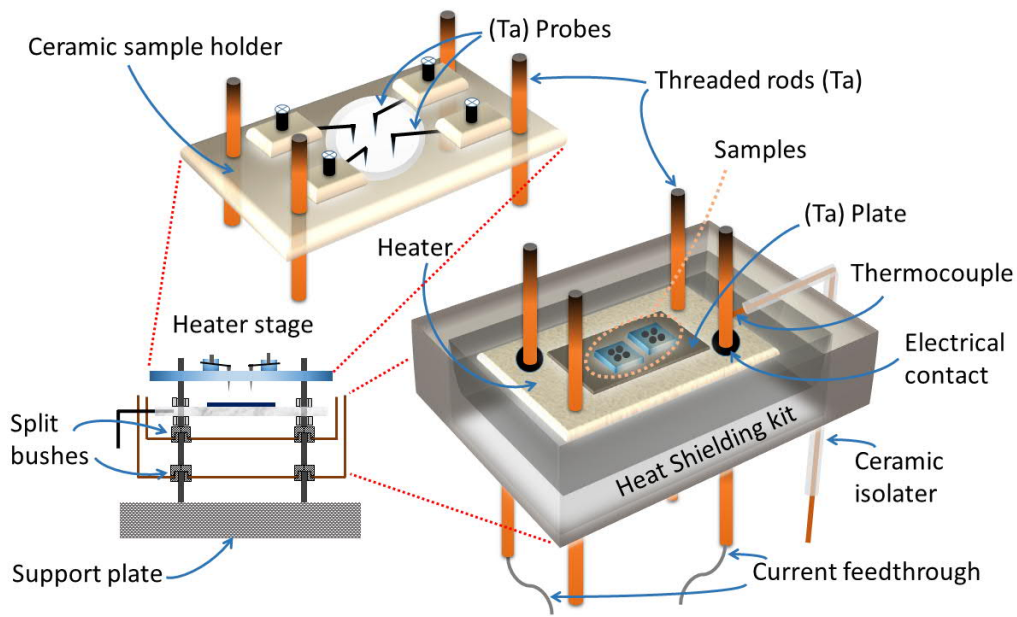

FIG. 1. The schematics of the ceramic sample holder and heater used in the system.

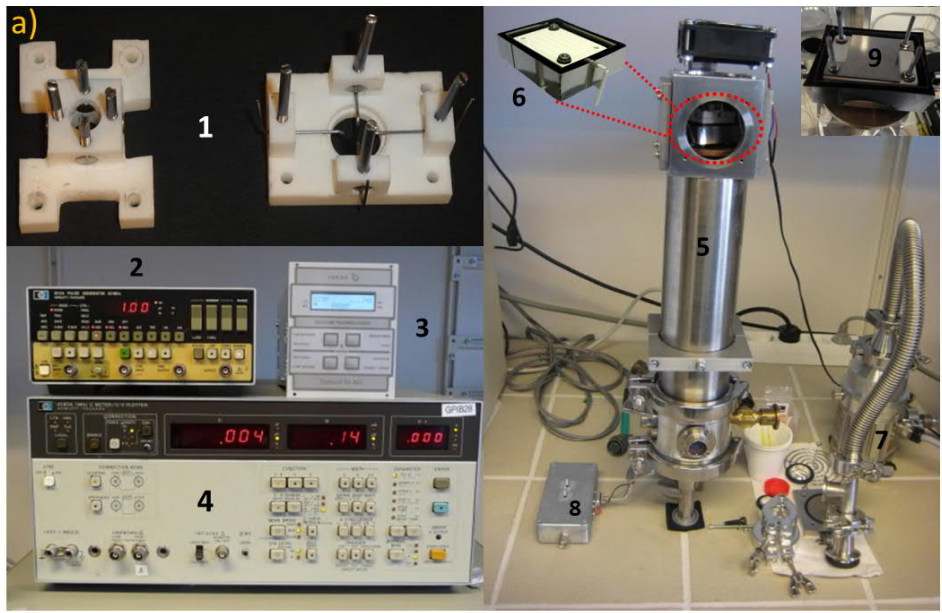

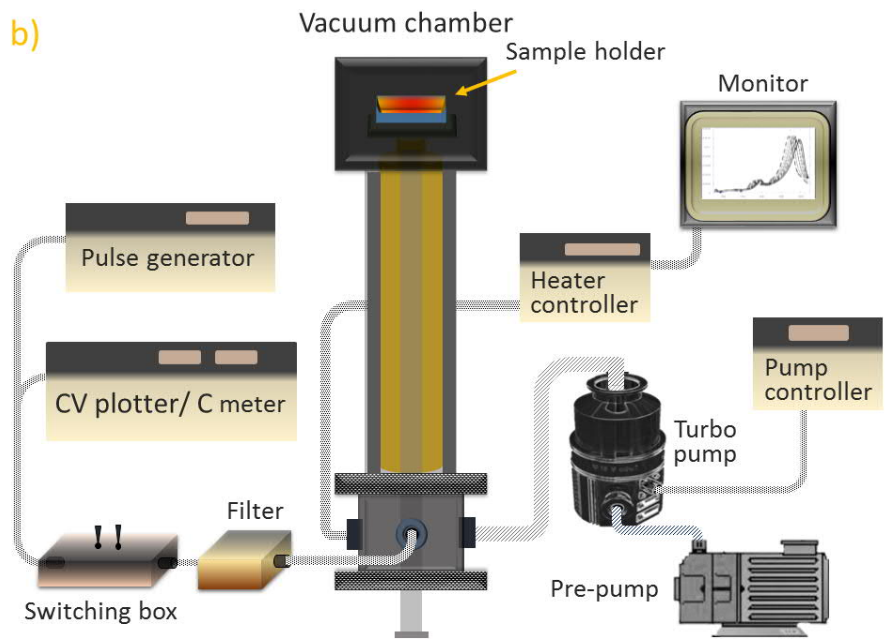

FIG. 2. Show a) the equipment used in the constructed DLTS. On the left, 1) ceramic sample holder, 2) HP 8112 pulse generator, 3) turbo pump controller and 4) HP 4280A 1MHz C meter/CV plotter. On the right; image of the 5) vacuum chamber, 6) Boralectric heater, 7) turbo pump, 8) the electronic switching box and 9) Tantalum plate. b) the schematic connection of the equipment in the DLTS setup.

Multiple electrical characterization is possible using four Ta probes mounted in an electrically insulating holder, made of the ceramic material, Macor (stable up to 1200 °C). Electrical measurements are performed by using a HP/Agilent 4280A 1 MHz (C meter / CV plotter) which can record transient signals needed for extracting high resolution DLTS signals coupled with a HP/Agilent 50 MHz 8112A programmable dual pulse generator, controlled by a GPIB card. To suppress the background noise a low pass filter has been built. A manual switch box was constructed to enable measurements on different sample configurations. Running and monitoring the experiment is fully automated by LabVIEW software.

The temperature range operation of the DLTS system was investigated by performing a series of heating test using a synthetic diamond sample. The sample was mounted in the ceramic sample holder, heated up to 1100 K in vacuum and remained in such condition for an hour. This procedure was repeated and successfully carried out with no evidence of degradation on neither sample nor the parts in the system.

**Experimental**

In the present work, an *n*-type 4H-SiC sample was prepared and analyzed for test and calibration of the setup. The 4H-SiC Schottky diodes were manufactured from *n*-type substrate wafers with a 10 µm thick *n*-type epitaxial layer purchased from Cree. The Schottky barrier is provided by circular Ni (150 nm) contacts, produced by electron beam evaporation. 200 nm Al was deposited on the backside of the sample whereas the ohmic contact was achieved using conducting silver paint to the highly doped substrate.

To ensure good rectification, the Schottky diodes were characterized by current-voltage (IV) and capacitance-voltage (CV) measurements determining the doping concentrations and overall functionality. The CV measurements were performed from room-temperature up to the breakdown point of the electrodes at 950 K (see Fig. 3). Typical net doping concentrations, $N_d - N_a$ determined from the slope of the $1/C^2$ vs reverse voltage, were in the range of $2 \times 10^{15}$ cm$^{-3}$ at room-temperature.

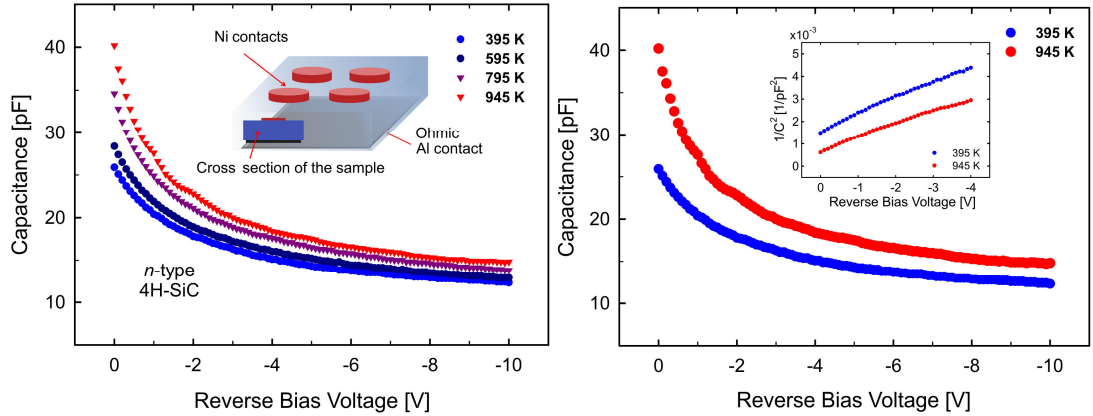

FIG. 3. Capacitance voltage of the reverse biased junction in *n*-type 4H-SiC, measured at different temperatures. The insets describe; (left) the schematic of the diode and (right) $1/C^2$ characteristic of the 4H-SiC sample.

For observation of the intrinsic deep-levels in *n*-doped 4H-SiC, HT-DLTS was performed in the temperature range 300 to 950 K. The following DLTS parameters were used: filling pulse width of 50 ms and a filling pulse height of 10 V together with the steady reverse bias $V_r = -10$ V. To evaluate the capacitance transient the conventional lock-in amplification were used to reduce the noise level and to achieve a high signal to noise ratio. The signal is obtained by pulsing a change in applied bias, and observing a transient decay of trapped charge carriers in the depletion region. The analysis of the capacitance transients follows a scheme outlined in [Ref. 18], but where arbitrary wave functions can be used for the evaluation.

**Results and Discussions**

DLTS studies are highly sensitive measurements, retaining certain demands on the samples. The maximum concentration of defects should preferably not exceed 10% of the doping density[19] to enable reliable quantitative analysis. For WBG materials, very high temperatures are needed to access the entire bandgap. The capacitance transient is studied as a function of temperature and, after signal processing, one obtains the DLTS spectrum. An electrically active deep-level appears in the spectrum as a peak for a certain temperature and measurement frequency. The peak height is directly proportional to the (uniform) trap concentration. Analyzing the peak positions provides the information about energy position in the bandgap and the carrier capture cross section. The analytic method used in our experiment is based on the conventional lock-in principal[18].

In high-quality *n*-type 4H-SiC epilayers, two deep-levels are often seen in DLTS positioned at 0.56-0.71 eV and 1.55-1.65 eV below the conduction band edge. These levels have now been clearly shown to originate from different charge states of the carbon vacancy[6]. The more shallow state, $Z_{1/2}$, is a double acceptor level, where the first transition is (-2|-1) is very fast and normally only the second transition is seen in DLTS (-1|0). The deeper state is named $EH_{6/7}$, but it is only $EH_7$ of this complex that belongs to the carbon vacancy and the donor transition (0|+1).

Here, the DLTS peaks of the $Z_{1/2}$ and the $EH_{6/7}$ transitions can clearly be observed in the sample measured on a Schottky structure (see Fig 4.). The activation energies are found by measuring a set of emission rates ($e_n$) as function of temperature and plotting $\ln(e_n/T^2)$ vs $1/T$, often referred to as an Arrhenius plot. The results achieved from the slope of the Arrhenius plot revealed full agreement with the data in the literatures, 0.68 eV and 1.55 eV below the conduction band for $Z_{1/2}$ and $EH_{6/7}$ respectively.

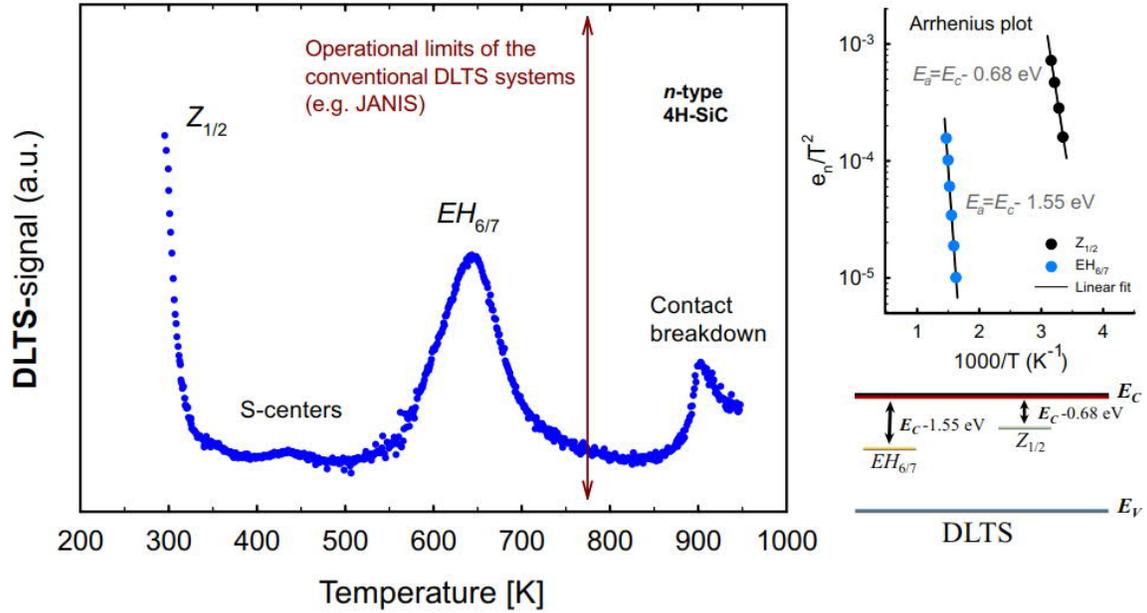

FIG. 4. HT-DLTS spectrum measured with a period width of 50 ms in the temperature range from 300 to 950 K obtained from $n$-typed 4H-SiC Schottky structure. Arrhenius plot gives the related position of the defects in the bandgap.

The chamber pressure was at a constant level of $5 \times 10^{-6}$ mbar during the entire measurement after 30 min of pre-pumping. Having a low pressure is important to extend the lifetime of the electrodes and increase the operational limit of the Schottky diodes up to higher temperatures. In DLTS measurements, it is crucial for detection of the small change in the capacitance to achieve a stable rectification. The capacitance transient is highly sensitive to the fluctuation of the conductivity. Hence, when the conductivity increases at elevated temperatures due to the excess of higher thermionic energy the emission process becomes much faster which reduce the rectification, increase the leakage current and conductance and finally results in breakdown of the electrodes. In Fig 5, one can see how the conductivity at a certain point, rapidly increases and reaches a critical level. From this point it is no longer possible to detect a DLTS signal. This is most likely due to the degradation and finally failure of the Ni contacts Schottky properties. According to earlier studies, at temperatures around 600 °C, $Ni_2Si$ phase formation appeals at the Ni/SiC interface. The transition from Schottky to ohmic contact in nickel silicide/SiC system during annealing is a well-known phenomenon.[20,21]

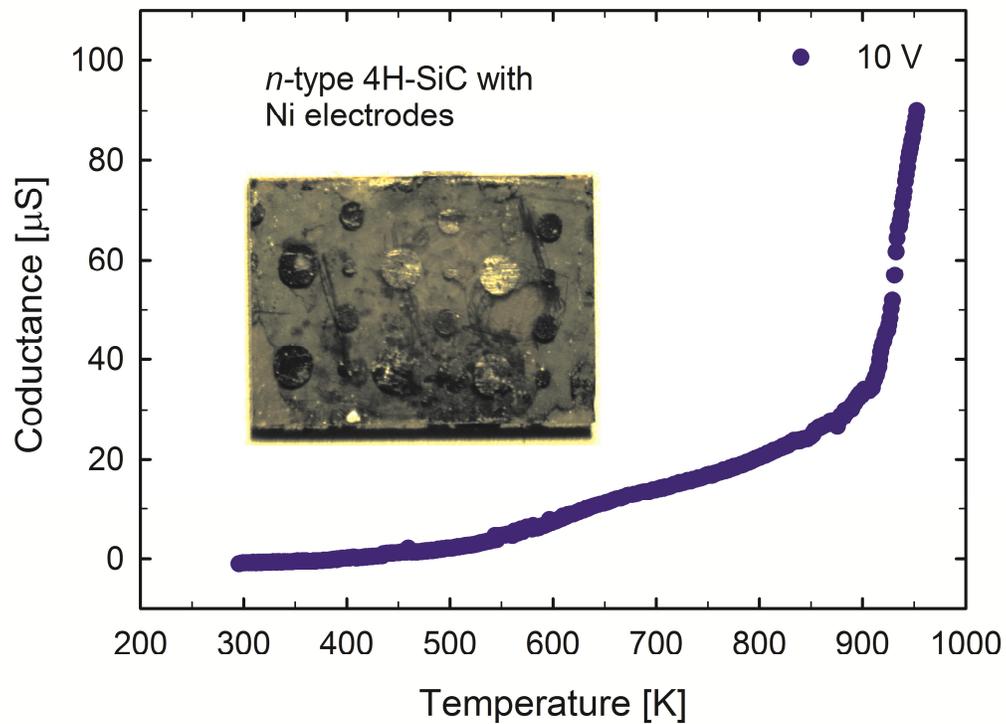

FIG. 5. The change of the conductance in *n*-type 4H-SiC sample are recorded as function of temperature at 10 V applied bias.

**Conclusions**

In summary, we have developed a high-temperature deep-level transient spectroscopy system for investigating deep defects in WBG semiconductor materials. Many of the important features of the defects and trap centers in such materials have not been identified so far. A reason is the lack of suitable electrical and high temperature stable environment in conventional DLTS setups. However, with the higher temperature range of this system, it is now possible to explore the bandgap of semiconductors much deeper than before.

Here, we provide experimental evidence by using an *n*-type 4H-SiC sample for calibration and to determine the function of the built setup. DLTS measurements were successfully performed up to 950 K where the well-known defect complex *$EH_{6/7}$* could be observed and the energy position in the bandgap extracted.

Though, the DLTS measurements could not reach up to the temperature limit of the system at 1100 K due to the failure of the Schottky contacts. Hence, achieving high quality diodes with good rectifying characteristics for operation at elevated temperatures is extremely important to fully use the potential of the system.

**Acknowledgements**

The Carl Trygger Foundation (13:284), LM Ericsson Research Foundation and Magnus Bergvalls Foundation are gratefully acknowledged for financial support.